\documentclass[twocolumn, prl, superscriptaddress]{revtex4}
\usepackage{graphicx,color,hyperref}
\usepackage{dcolumn}
\usepackage{bm}
\usepackage{mathrsfs}
\usepackage{amsfonts}
\usepackage{amsmath}
\usepackage{amssymb}
\usepackage{graphics}
\usepackage{subfigure}
\usepackage{graphicx}
\usepackage{threeparttable}
\usepackage{booktabs}
\usepackage{tabularx}
\usepackage{amsmath}

\begin{document}

\title{\textit{SU(2)} Ginzburg-Landau theory for degenerate Fermi gases with
synthetic non-Abelian gauge fields}
\author{Kuang Zhang}
\affiliation{State Key Laboratory of Quantum Optics and Quantum Optics Devices, Institute
of Laser spectroscopy, Shanxi University, Taiyuan 030006, P. R. China}
\author{Yanlin Feng}
\affiliation{State Key Laboratory of Quantum Optics and Quantum Optics Devices, Institute
of Laser spectroscopy, Shanxi University, Taiyuan 030006, P. R. China}
\author{Chuanwei Zhang}
\affiliation{Department of Physics, the University of Texas at Dallas, Richardson, TX
75080 USA}
\author{Gang Chen}
\affiliation{State Key Laboratory of Quantum Optics and Quantum Optics Devices, Institute
of Laser spectroscopy, Shanxi University, Taiyuan 030006, P. R. China}
\author{Suotang Jia}
\affiliation{State Key Laboratory of Quantum Optics and Quantum Optics Devices, Institute
of Laser spectroscopy, Shanxi University, Taiyuan 030006, P. R. China}

\begin{abstract}
\textbf{The non-Abelian gauge fields play a key role in achieving novel
quantum phenomena in condensed-matter and high-energy physics. Recently, the
synthetic non-Abelian gauge fields have been created in the neutral
degenerate Fermi gases, and moreover, generate many exotic effects. All the
previous predictions can be well understood by the microscopic
Bardeen-Cooper-Schrieffer theory. In this work, we establish an \textit{SU(2)%
} Ginzburg-Landau theory for degenerate Fermi gases with the synthetic
non-Abelian gauge fields. We firstly address a fundamental problem how the
non-Abelian gauge fields, imposing originally on the Fermi atoms, affect the
pairing field with no extra electric charge by a local gauge-field theory,
and then obtain the first and second \textit{SU(2)} Ginzburg-Landau
equations. Based on these obtained \textit{SU(2)} Ginzburg-Landau equations,
we find that the superfluid critical temperature of the intra- (inter-) band
pairing increases (decreases) linearly, when increasing the strength of the
synthetic non-Abelian gauge fields. More importantly, we predict a novel
\textit{SU(2)} non-Abelian Josephson effect, which can be used to design a
new atomic superconducting quantum interference device.}
\end{abstract}

\maketitle

%\pacs{67.85.Lm, 74.20.Rp, 67.85.Lm, 74.20.Fg}

The non-Abelian gauge fields, whose different components do not commute each
other, are a central building block of the theory of fundamental
interactions. Attributed to their high degrees of controllability,
tunability, and versatility, ultracold quantum gases are a powerful platform
to simulate the non-Abelian gauge fields. In general, the atomic quantum
gases are charge neutral, and are thus not influenced by external gauge
fields the way electrons are. Fortunately, by controlling different
laser-atom interactions, the synthetic non-Abelian gauge fields can be
created in these neutral quantum gases \cite{RJ95,JD11,NG14}. Moreover, the
simplest non-Abelian gauge field, which is always called the one-dimensional
(1D) equal-Rashba-Dresselhaus(ERD)-type spin-orbit coupling, has been
realized experimentally \cite%
{YJL11,JYZ12,CQ13,JSC14,CH14,PW12,RAW13,ZF14,LWC12}, using a pair of Raman
lasers. Recently, the similar but spatial-dependent gauge field has also
been achieved in ultracold $^{87}$Rb atom \cite{MCB13}. These important
experiments pave a new way for exploring nontrivial quantum effects, induced
by the synthetic non-Abelian gauge fields, in ultracold quantum gases. For
instance, based on the microscopic Bardeen-Cooper-Schrieffer (BCS) theory,
exotic superfluids \cite%
{RYW12,YZQ11,HH11,VJP11,HL12,WF13L,WF13A,YX14,MG11,MG12,KS12,HH13,CC13,CQU13,WZ13, XJL13,CCF14,HH14}%
, including the topological BCS \cite{MG11,MG12,KS12,HH13,CC13} and
Fulde-Ferrell-Larkin-Ovchinnikov phases \cite{CQU13, WZ13, XJL13,CCF14,HH14}%
, have been predicted in degenerate Fermi gases.

In the conventional charge superconductors, the \textit{U(1)}
Ginzburg-Landau (GL) theory, in parallel with the microscopic BCS theory, is
another famous theory to explore relevant physics \cite{MC73}. One of its
most powerful features that it can be used to quantitatively describe the
effects induced thermal fluctuations in the intermediate and strong coupling
normal states, which are, however, missed in the BCS theory \cite{CAR93}.
Moreover, some novel quantum phenomena, such as Josephson effect, flux flow,
and the melting of the Abrikosov vortex lattice, \emph{etc.} \cite{JFA04},
have also been revealed by this theory. However, the GL theory for
degenerate Fermi gases with the synthetic non-Abelian gauge fields is still
lacking. In this work, we establish an \textit{SU(2)} GL theory for this
system, based on the non-Abelian properties of the synthetic gauge fields.

Notice that in the conventional charge superconductors, the pairing has the
electric charge $2e$, and is thus affected easily by the external gauge
fields. However, the formed pairing in degenerate Fermi gases is charge
neutral. It is natural to ask a fundamental and very important problem how
the neutral pairing field interacts with the synthetic non-Abelian gauge
fields, imposing originally on the Fermi atoms. We firstly address this key
issue by a local gauge-field theory of the pairing field. Then, we obtain
the first and second \textit{SU(2)} GL equations by the variation of the
total free energy with respect to the pairing field and the synthetic
non-Abelian gauge fields. Based on these obtained \textit{SU(2)}
GL equations, we find that the superfluid critical temperature
of the intra- (inter-) band pairing increases (decreases) linearly, when
increasing the strength of the synthetic non-Abelian gauge fields. More
importantly, we predict a novel \textit{SU(2)} non-Abelian Josephson effect,
which can be used to design a new atomic superconducting quantum
interference device.

\bigskip

{\LARGE \textbf{Results}}

\textbf{Total free energy in space.} In general, the pairing field,
resulting from the two-component Fermi atom field $\phi (\mathbf{r})$
coupled with the synthetic \textit{SU(2)} non-Abelian gauge fields, is
expressed as $\psi (\mathbf{R})=\phi _{1}(\mathbf{r})\phi _{2}(\mathbf{r}%
^{\prime })$, where $\phi _{1}$ and $\phi _{2}$ are the fields for two
different Fermi atoms, $\mathbf{r}$ is the 3D space-dependent coordinate of
the Fermi atom, and $\mathbf{R}=(\mathbf{r}+\mathbf{r}^{\prime })/2$ is the
coordinate of the pairing field \cite{AJL06}. Obviously, $\psi $ is a boson
field. In terms of the local gauge-field theory of the pairing field (see
Methods), we demonstrate strictly that this pairing field has an internal
helical doublet and can interact with the same synthetic non-Abelian gauge
fields, imposing originally on the Fermi atoms.

In addition, the total free energy is derived, in units of $\hbar =c=1$, by
(see Methods)
\begin{equation}
F_{\text{s}}=\int d^{3}\mathbf{R(}f_{\text{n}}+U_{\text{eff}}+f_{\text{c}%
}+f_{\text{G}}).  \label{FEF}
\end{equation}%
In equation (\ref{FEF}), $f_{\text{n}}$ is the energy density of the normal
state. $U_{\text{eff}}=a\psi ^{\ast }\psi +b(\psi ^{\ast }\psi )^{2}/2$,
with the coefficients $a$ and $b$, is the effective potential of the pairing
field. The first and second terms of $U_{\text{eff}}$ are the free and
self-interacting energy densities of the pairing field, respectively. The
explicit expressions of the coefficients $a$ and $b$ can, in principle, be
determined from the microscopic BCS theory \cite{AJL06}. In general, the
coefficient $a$ is dependent of temperature. When the temperature is lower
than the superfluid critical temperature, $a>0$, while $a<0$ vice versa. On
contrary, the coefficient $b$ is positive for any temperature. $f_{\text{c}%
}=\left\vert \Pi _{i}\psi \right\vert ^{2}/4m$ is the kinetic energy, where $%
\Pi _{i}=-i\partial _{i}-\alpha A_{i}$ and $m$ is the mass of the Fermi
atom. $f_{\text{G}}=\varrho L_{ij}L_{ij}$ is the energy density functional
of the synthetic non-Abelian gauge fields $A_{i}$, where $L_{ij}=\Pi
_{i}A_{j}-\Pi _{j}A_{i}$ is the tensor of the synthetic non-Abelian gauge
fields $A_{i}$, and satisfies the anti-symmetry property $L_{ij}=-L_{ji}$,
and $\varrho $ is a coefficient determined by the synthetic
non-Abelian gauge fields $A_{i}$. In the previous discussions, the synthetic
non-Abelian gauge fields are usually chosen, in the spin-basis representation, as
\begin{equation}
\mathbf{A}=[l\sigma _{x},\chi l\sigma _{y},0,f(l)\text{\^{I}}_{2}\mathbf{]},
\label{GFE}
\end{equation}%
where $l$ and $f(l)$ are the introduced functions of space-time, the
dimensionless constant $\chi $ determines the type of the synthetic
non-Abelian gauge fields, and \^{I}$_{2}$ is a $2\times 2$ unit matrix. For $%
\chi =1$, the 2D RD-type non-Abelian gauge field emerges, and becomes the 1D
ERD-type non-Abelian gauge field in the case of $\chi =0$ \cite%
{YJL11,JYZ12,CQ13,JSC14,CH14,PW12,RAW13,ZF14,LWC12}. Recent experiment shows
that the functions $l$ and $f(l)$ can be determined by the Rabi frequencies
of laser fields \cite{MCB13}, thus both space- and time- dependent functions
$l$ and $f(l)$ can be accessible.

\textbf{The first }\textit{SU(2)} \textbf{GL equation.} To describe the
stable superfluid, we need study the variations of the total free energy, $%
\delta F_{\text{s}}(\psi )$, $\delta F_{\text{s}}(\psi ^{\ast }) $, and $%
\delta F_{\text{s}}(A_{i})$. In the case of the three-component non-Abelian
gauge fields $A_{i}$, the results are very complicate. For simplicity, here
we only deal with the in-plane non-Abelian gauge fields, i.e., $A_{z}=0$. In
such a case, we obtain the first \textit{SU(2)} GL equation (see Methods)
\begin{equation}
\frac{1}{4m}\left[ (-i\partial _{\zeta }-\alpha A_{\zeta })^{2}-\partial
_{z}^{2}\right] \psi +a\psi +b\psi ^{2}\psi =0  \label{GL1}
\end{equation}%
with $\zeta =x,y$.

The gauge-invariant field equation (\ref{GL1}) fully describes the interplay
between neutral superfluids and the synthetic non-Abelian gauge fields, when
the temperature is lower than the superfluid critical temperature. It seems
that this equation is similar to that of the \textit{U(1)} case. In fact,
the physics is quite different. Attributed to the \textit{SU(2)} properties
of the synthetic gauge fields, there are two kinds of superfluid
states, including the positive and negative helical states. Moreover, they
couple with each other and both of them are vectors in 2D Hilbert space of
the helical basis. It means that equation (\ref{GL1}) is a two-component
coupled equation in 2D Hilbert space. In addition, in the \textit{U(1)}
case, the pairing is formed by two spin states. However, in the presence of
the synthetic non-Abelian gauge fields, the pairing emerges in two helical
states. These different spin and helical states lead to different dispersion
relations, and thus different microscopic quantum statistics of the
interacting many-body systems. It implies that the coefficients $a$ and $b$
are also different.

Due to existence of the term $b\psi ^{2}\psi $, the two-component nonlinear
equation (\ref{GL1}) is hard to be solved exactly. Here we use an
approximate linearization method (i.e., assuming $b\psi ^{2}\psi \simeq 0$)
to deal with this equation \cite{AAA57,BR10}. As an example, we consider a
static RD-type non-Abelian gauge field, i.e., $l=1$ and $\chi =1$ in
equation (\ref{GFE}). In such case, we rewrite the spatial part of this
non-Abelian gauge field as $[\sigma _{x}k_{\text{F}}(\xi _{0}+\varkappa
y),\sigma _{z}k_{\text{F}}(\xi _{0}+\varkappa x),0\mathbf{]}$, with the
dimensionless infinitesimal $\varkappa $ and the Fermi vector $k_{\text{F}}$
of the non-interacting Fermi gases, and then assume the corresponding
solution as $\psi =\exp (ik_{z}z)h(x,y)$. The introduced dimensionless
infinitesimal $\varkappa $ doesn't change the static property of the RD-type
non-Abelian gauge field since $\varkappa y\longrightarrow 0$ and $\varkappa
x\rightarrow 0$, but is an auxiliary quality, which only help us to approximately
solve equation (\ref{GL1}). Substituting the assumed solution $\psi $ into
equation (\ref{GL1}) and using the approximate linearization method \cite%
{AAA57,BR10}, we obtain the following 2D oscillator-type equation: $%
-(\partial _{x}^{2}+\partial _{y}^{2})h(x,y)/4m+m\varkappa ^{2}[\omega
_{cy}^{2}(y-C_{0})^{2}+\omega _{cx}^{2}(x-C_{0})^{2}]h(x,y)=(\left\vert
a\right\vert -k_{z}^{2}/4m)h(x,y)$, where $\omega _{cx}=\alpha \sigma _{z}k_{%
\text{F}}/2m$ and $\omega _{cy}=\alpha \sigma _{x}k_{\text{F}}/2m$ are the
circular frequencies in the $x$ and $y$ directions, respectively, and $%
C_{0}=-\xi _{0}/\varkappa $. By further solving the above oscillator-type
equation, we obtain $\left\vert a\right\vert -k_{z}^{2}/4m$ $=(n_{x}^{\prime
}+1/2)\omega _{cx}+(n_{y}^{\prime }+1/2)\omega _{cy}$, where $n_{x}^{\prime
} $ and $n_{y}^{\prime }$ are the positive integers. When the condensate
occurs, only the ground state ($n_{x}^{\prime }=n_{y}^{\prime }=0$, $k_{z}=0$%
) becomes significant \cite{AAA57,BR10}. As a consequence, the critical
temperature is obtained, in the spin-basis representation, by $T_{c}^{\text{s%
}}\simeq T_{c}(0)-\alpha k_{\text{F}}(\sigma _{x}+\sigma _{z})/4a_{T}m $,
where $T_{c}(0)$ is the critical temperature without the synthetic
non-Abelian gauge fields, and $a_{T}(>0)$ is the leading-order expansion
coefficient of $a$ at $T_{c}(0)$.

Since in this work we investigate the physics of superfluid with the helical
doublet, the critical temperature is obtained, from a transformation of
\textit{SU(2)} group representation to the helical basis of pairing doublet,
by
\begin{equation}
T_{c}\simeq T_{c}(0)-\frac{\sqrt{2}\alpha k_{\text{F}}\sigma _{z}}{4a_{T}m}.
\label{SCT}
\end{equation}%
When $\alpha =0$, $T_{c}=T_{c}(0)$, as expected. Equation (\ref{SCT}) shows
that the superfluid critical temperature is a $2\times 2$ matrix, because
equation (\ref{GL1}) is a two-component coupled equation. The diagonal
elements reflect the critical temperature for the different superfluid
states (the positive and negative helical states). Using the similar
consideration of the electric charge matrix of the left-handed doublet of
lepton \cite{SW67}, we find that, when increasing the coupling strength $%
\alpha $, the critical temperature of the pairing field in the negative
helical state increases linearly from a non-zero value, which is consistent
with the result derived from the microscopic BCS theory with the Nozie\'{r}%
es--Schmitt-Rind correction \cite{RYW12}. Moreover, we can confirm that the
pairing fields in the positive and negative helical states govern the
superfluid physics of the inter- and intra- band pairings, respectively. For
the superfluid critical temperature of the inter-band pairing, it decreases
linearly when increasing the coupling strength $\alpha $. This behavior can
also be easily understood since the inter-band pairing is gradually
suppressed, attributed to the blocking effect in Fermi surface.

\textbf{The second \textit{SU(2)} GL equation.} The second \textit{SU(2)} GL
equation is obtained by (see Methods)
\begin{equation}
\frac{i\alpha }{4m}(\psi ^{\ast }\partial _{\zeta }\psi -\psi \partial
_{\zeta }\psi ^{\ast })+\frac{\alpha ^{2}}{2m}\psi ^{\ast }\psi A_{\zeta
}=2\varrho (\Lambda _{\zeta }-i2\alpha \Theta _{\zeta }),  \label{GL2}
\end{equation}%
where
\begin{equation}
\Lambda _{\zeta }=\partial _{\eta }\partial _{\zeta }A_{\eta }-\partial
_{\eta }^{2}A_{\zeta }-\partial _{z}^{2}A_{\zeta },  \label{GL2-S1}
\end{equation}%
\begin{equation}
\Theta _{\zeta }=-(\partial _{\zeta }A_{\eta })A_{\eta }+(\partial _{\eta
}A_{\zeta })A_{\eta }  \label{GL2-S2}
\end{equation}%
with $\eta =x,y$. Equation (\ref{GL2}) is also a two-component field
equation. The left term of this equation reflects the in-plane supercurrents
\cite{AJL06}, i.e,
\begin{equation}
j_{\zeta }=\frac{i\alpha }{4m}(\psi ^{\ast }\partial _{\zeta }\psi -\psi
\partial _{\zeta }\psi ^{\ast })+\frac{\alpha ^{2}}{2m}\psi ^{\ast }\psi
A_{\zeta },  \label{GL3}
\end{equation}%
This means that equation (\ref{GL2}) governs the interplay between the
in-plane supercurrents $j_{\zeta }$ and the synthetic non-Abelian gauge
fields $A_{\zeta }$. In addition, the term $\Theta _{\zeta }$ in equation (%
\ref{GL2-S2}) is a new term, originating from the non-Abelian properties of
the synthetic gauge fields $A_{\zeta }$. The supercurrent in the $z$\
direction is given by (see Methods)
\begin{equation}
j_{z}=\frac{i\alpha }{4m}(\psi ^{\ast }\partial _{z}\psi -\psi \partial
_{z}\psi ^{\ast }).  \label{R8}
\end{equation}

In terms of Noether's theorem \cite{LR96}, the neutral supercurrents in
equations (\ref{GL3}) and (\ref{R8}) are the \textit{SU(2)} charge currents,
rather than the conventional probability currents\textbf{\ }($%
j_{i}=n_{0}\partial _{i}\theta /2m$) of superfluid order parameter $\psi
_{0}=\sqrt{n_{0}}e^{i\theta }$\ without any gauge field, where $n_{0}$\ is
the density of pairing. However, the supercurrent in the $z$ direction is
trivial, since it doesn't interact with the synthetic non-Abelian gauge
fields $A_{\zeta }$. When the synthetic gauge fields are the \textit{U(1)}
cases, the term $\Theta _{\zeta }=0$, and equations (\ref{GL1}) and (\ref%
{GL2}) reduce respectively to $\left[ (-i\partial _{\zeta }-e^{\prime
}A_{\zeta })^{2}-\partial _{z}^{2}\right] \psi /4m+a\psi +b\psi ^{2}\psi =0$
and $ie^{\prime }(\psi ^{\ast }\partial _{\zeta }\psi -\psi \partial _{\zeta
}\psi ^{\ast })/4m+e^{\prime 2}\psi ^{\ast }\psi A_{\zeta }/2m=2\varrho
\Lambda _{\zeta }$, where $e^{\prime }$ is the effective electric charge
\cite{AJL06}. Moreover, the pairing field $\psi $ is a single-component
scalar field.

We emphasize that the nonlinear \textit{SU(2)} GL equation (\ref{GL2}) is
gauge invariant, even if the terms $\Lambda _{\zeta }\ $and $\Theta _{\zeta
} $ are dependent of the synthetic non-Abelian gauge fields. Notice that for
the static synthetic non-Abelian gauge fields, $\Lambda _{\zeta }=0$\ and $%
\Theta _{\zeta }=0$, derived from equations (\ref{GL2-S1})-(\ref{GL2-S2}).
It seems that the in-plane supercurrents vanish. In fact, in terms of
\textit{SU(2)} symmetry, we make a local gauge transformation $\psi ^{\prime
}=U_{L}\psi $\ and $A_{i}^{\prime }=-i(\partial _{i}U_{L})U_{L}^{\dag
}/\alpha +U_{L}A_{i}U_{L}^{\dag }$, and then obtain $\Lambda _{\zeta
}^{\prime }\neq 0$\ and $\Theta _{\zeta }^{\prime }\neq 0$, i.e., equation (%
\ref{GL2}) as well as the in-plane supercurrents\textbf{\ }still exist.

If the synthetic non-Abelian gauge fields are dependent of space-time, they
can not be transformed to the static cases by a local gauge transformation.
In this case, the superfluid physics becomes very rich, and however, is
difficult to be discussed by the microscopic BCS theory. On contrary, our
established GL theory is a powerful tool in this respect. In the Table I, we
give the explicit expressions of $\Lambda _{\zeta }$, $\Theta _{\zeta }$,
and especially, the supercurrents for the 2D RD- and 1D ERD- type
non-Abelian gauge fields with $l=\Sigma t+\varepsilon \sin (\omega
_{0}t)/\omega _{0}$, where the physical meanings of parameters $\Sigma $, $%
\varepsilon $, and $\omega _{0}$ will be interpreted in the following
discussions. For the 1D ERD-type non-Abelian gauge field, the new term $%
\Theta _{\zeta }$ disappears. In addition, we will show in the next section
that the space-time-dependent non-Abelian gauge fields generate a novel
\textit{SU(2)} non-Abelian Josephson effect, which is a tunneling phenomenon
in a weakly-linked superfluid system \cite{BD65}.

\begin{figure}[t]
\centering
\includegraphics[width=8.0cm,height=5.50cm]{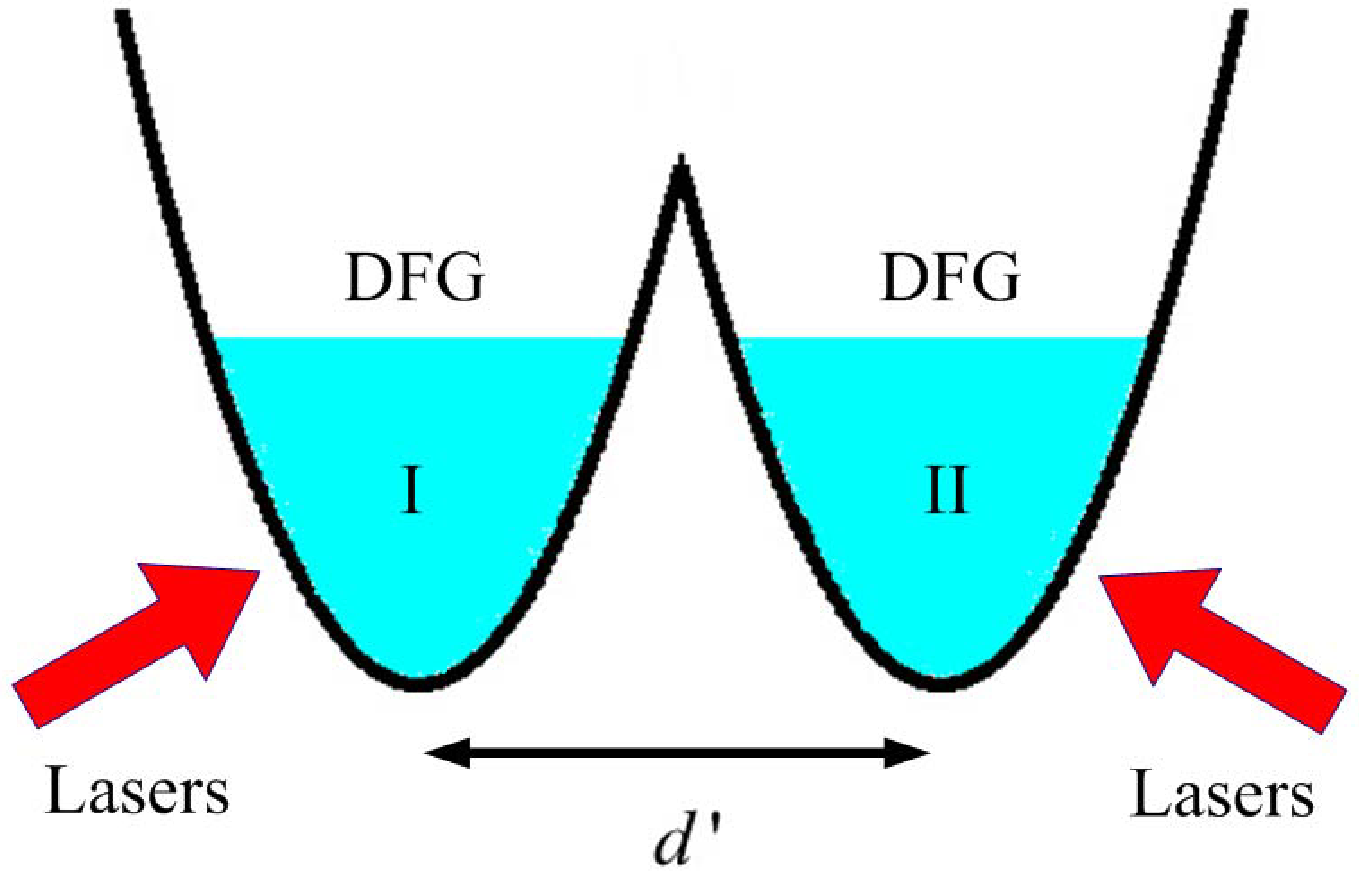}\newline
\caption{\textbf{A possible scheme to achieve the \textit{SU(2)} non-Abelian
Josephson effect and the Shapiro step.} Two identical degenerate Fermi gases
(DFGs) without initial population imbalance are respectively distributed in two sides of
the double-well potential. The lasers are used to create the non-Abelian
gauge fields.}
\label{fig1}
\end{figure}

\textbf{\textit{SU(2)} non-Abelian Josephson effect.} To predict the \textit{%
SU(2)} non-Abelian Josephson effect, we consider two identical degenerate
Fermi gases without initial population imbalance, respectively distributed
in two sides of double-well potential through a weakly-linked barrier \cite%
{LS07,AM05,SA07,HH111}, and denote these two regions as I and II (see Fig.%
\ref{fig1}). The space-time-dependent non-Abelian gauge fields are chosen as
the terms in equation (\ref{GFE}). In this neutral Fermi atom system, we can
investigate a gauge-invariant mass current $j_{\zeta }^{m}=2mj_{\zeta }$,
where the factor $2$ originates that the formed pairing consists of two atoms
\cite{BPA98,FSC01}.
  
\renewcommand{\arraystretch}{1.5}
\begin{center}
\begin{threeparttable}[t]
\caption{ \textbf{The explicit expressions of $\Lambda _{\zeta }$, $\Theta _{\zeta }$, and especially, the supercurrents for the 2D RD- and 1D ERD- type non-Abelian gauge fields}.
The supercurrent 1 can be influenced by the synthetic non-Abelian gauge fields,
while the supercurrent 2 is only a trivial \textit{SU(2)} charge current, like Eq. (\ref{R8}).
In the case of the 1D ERD-type non-Abelian gauge field, the new term $\Theta _{\zeta }$ disappears,
and the supercurrent 1 emerges only in the $x$ direction. Here, the functions are defined as $y_{1}=\Sigma t^{2}\partial _{\eta }\Sigma +t\partial _{\eta }(\Sigma \varepsilon
)\sin(\omega _{0}t)/\omega _{0}$ and $y_{2}=\varepsilon \partial _{\eta }\varepsilon \sin
^{2}(\omega _{0}t)/\omega _{0}^{2}$, respectively. The indices $\zeta$ and $\eta$ satisfy the Einstein rule, in which $\zeta$ and $\eta$ only take different coordinates of $x$ and $y$ at the same time, i.e., if $\zeta=x$, then $\eta=y$.}
\begin{tabularx}{85mm}{ccc}
%\begin{tabular}{ccc}
\toprule[2.2pt]
\specialrule{0.4pt}{3pt}{2pt}
 & ERD & RD \\ \hline
\textit{SU(2)} gauge fields & $(l\sigma_{x}, 0, 0)$ & $(l\sigma _{x}, l\sigma _{y}, 0)$
 \\
${\Lambda}_{\zeta }$ & $-(\partial _{y}^{2}+\partial _{z}^{2})l\sigma _{x}$ & $-(\partial _{\eta}^{2}+\partial _{z}^{2})l\sigma _{\zeta }$  \\
$\Theta_{\zeta }$ & $0$ & $(y_{1}+y_{2})(\sigma_{\zeta}\sigma_{\eta}-\hat{I})$ \\
Supercurrent 1 & $j_{x}^{\text{ERD}}$ & $j_{x}^{\text{RD}}$, $j_{y}^{\text{RD}}$ \\
Supercurrent 2 & $j_{y}^{\text{ERD}}$, $j_{z}^{\text{ERD}}$ & $j_{z}^{\text{RD}}$ \\
\specialrule{0.4pt}{2pt}{1.5pt}
\bottomrule[1.4pt]
\end{tabularx}
%\end{tabular}
\end{threeparttable}
\end{center}

When the vacuum condensate occurs, we have two conditions, $\partial U_{%
\text{eff}}/\partial \psi =0$ and $\partial U_{\text{eff}}/\partial \psi
^{\ast }=0$ \cite{LR96}, and thus $\psi ^{\ast }\psi =-a/b$. So the pairing
field is written as $\psi =(\sqrt{a/4b},\sqrt{a/4b})^{T}e^{i(\frac{\pi }{2}%
+\varphi )}$, where $\varphi $ is the phase. In addition, in this
weakly-linked system, we also have two phenomenal boundary conditions, $%
\partial \psi _{\text{I}}/\partial \zeta =\psi _{\text{II}}/d^{\prime }$ and
its complex conjugate, at the barrier \cite{MC73}, where the parameter $%
d^{\prime }$ is the width of barrier. Without the synthetic non-Abelian
gauge fields $A_{\zeta }$, the direct-current Josephson mass current density
in the $\zeta $ direction is found as $j_{\zeta }^{m}=-(a\alpha /2bd^{\prime
})\hat{\Xi}\sin (\Delta \varphi )$, where the phase difference is defined as
$\Delta \varphi \mathbf{=}\varphi _{\text{II}}-\varphi _{\text{I}}$ and $%
\hat{\Xi}$ is a $2\times 2$ matrix with $\hat{\Xi}_{11}=\hat{\Xi}_{12}=\hat{%
\Xi}_{21}=\hat{\Xi}_{22}=1$. In the presence of the synthetic non-Abelian
gauge fields $A_{\zeta }$, the phase difference must be modified, in order
to obtain the gauge-invariant mass current. By considering the dimensional
property, we write a gauge-invariant phase difference as $\Delta \varphi
=\varphi _{\text{II}0}-\varphi _{\text{I}0}-\alpha \int\nolimits_{\text{I}}^{%
\text{II}}A_{\zeta }d\zeta $, where $\Delta \Phi _{0}=\varphi _{\text{II}%
0}-\varphi _{\text{I}0} $ is the initial phase difference between the
regions I and II. Since the synthetic non-Abelian gauge fields $A_{\zeta }$
are dependent of space-time, we further rewrite the total phase difference
as $\Delta \mathbf{\varphi =}-\alpha \int_{t_{0}}^{t}dt\int\nolimits_{\text{I%
}}^{\text{II}}\frac{dA_{\zeta }}{dt}d\zeta $ , and the mass current is thus
given by
\begin{equation}
j_{\zeta }^{m}=-\frac{a\alpha }{2bd^{\prime }}\left[
\begin{array}{cc}
1 & 1 \\
1 & 1%
\end{array}%
\right] \sin \mathbf{(-}\alpha \mathbf{\left[
\begin{array}{cc}
0 & \kappa _{1} \\
\kappa _{2} & 0%
\end{array}%
\right] ),}  \label{AC1}
\end{equation}%
where $\kappa _{1}=\int_{t_{0}}^{t}dt\int\nolimits_{\text{I}}^{\text{II}%
}\lambda _{1}\frac{dl}{dt}d\zeta $, $\kappa
_{2}=\int_{t_{0}}^{t}dt\int\nolimits_{\text{I}}^{\text{II}}\lambda _{2}\frac{%
dl}{dt}d\zeta $, and $\lambda _{1}$ and $\lambda _{2}$ are the dimensionless
constants, determined by the type of the synthetic non-Abelian gauge fields.
For the ERD-type ($\chi =0$) or the $x$ component of the RD-type ($\chi =1$)
non-Abelian gauge fields, $\lambda _{1}=\lambda _{2}=1$, which become $%
\lambda _{1}=-i$ and $\lambda _{2}=i$ in the case of the $y$ component of
the RD-type non-Abelian gauge field.

By controlling different laser-atom interactions (for example, adding a
sinusoidal perturbation on Rabi frequencies, etc.) \cite{RJ95,JD11,NG14}, we
can choose $dl/dt=\Sigma +\varepsilon \cos (\omega _{0}t)$ with $\Sigma
=dG/d\zeta $ and $\varepsilon =dg/d\zeta $, where $G$ and $g$ reflect the
chemical potential difference between the two wells of the unit \textit{SU(2)%
} charge in the synthetic non-Abelian gauge fields and its amplitude of
oscillating potential perturbation, respectively. In this case, the
coefficient $\varrho =\omega _{0}d^{\prime 4}/4$. When $\{G,g\}\ll E_{\text{F%
}}$ ($E_{\text{F}}=k_{\text{F}}^{2}/2m$ is the Fermi energy of the
non-interacting gases), equation (\ref{AC1}) is simplified as (see Methods)
\begin{equation}
j_{\zeta }^{m}=-\frac{a\alpha }{2bd^{\prime }}\hat{\Xi}\sum\limits_{k=-%
\infty }^{\infty }\hat{\Omega}_{k}  \label{AC2}
\end{equation}%
with $\hat{\Omega}_{k}=$diag$(\kappa _{3},\kappa _{4})$, where $\kappa
_{3}=\lambda _{1}\lambda _{2}(-1)^{k}J_{k}(\alpha g/\omega _{0})\sin
[(\alpha G-k\omega _{0})t+\alpha \Delta \varphi _{01})]$, $\kappa
_{4}=\lambda _{1}\lambda _{2}(-1)^{k}J_{k}(\alpha g/\omega _{0})\sin
[(\alpha G-k\omega _{0})t+\alpha \Delta \varphi _{02})]$, $J_{k}(\alpha
g/\omega _{0})$ is the $k$-th Bessel function with respect to $\alpha
g/\omega _{0}$, and $\Delta \varphi _{01}$ and $\Delta \varphi _{02}$ are
the matrix elements of the initial phase difference. Equation (\ref{AC2})
shows that the space-time-dependent synthetic non-Abelian gauge fields can
induce an alternating-current \textit{SU(2) }Josephson mass current, which
has never been predicted from the microscopic BCS theory \cite%
{YZQ11,HH11,VJP11,HL12,RYW12,WF13L,WF13A,YX14,MG11,MG12,KS12,HH13,CC13,CQU13,WZ13, XJL13,CCF14,HH14}%
.

\begin{figure}[t]
\centering
\includegraphics[width=6.50cm,height=5.50cm]{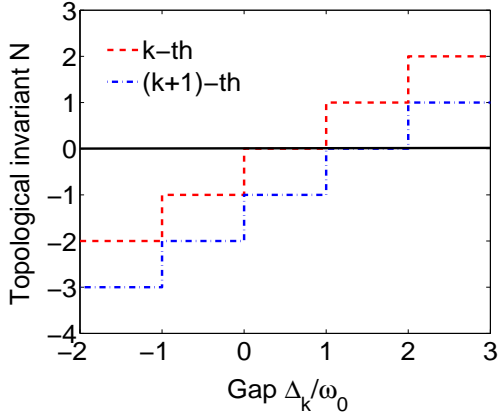}\newline
\caption{\textbf{The evolution of the topological invariant of the current
component in equation (\protect\ref{AC2}).} The black line represents the
topologically-trivial mass current in the presence of a static Rashba-type
non-Abelian gauge field.}
\label{fig2}
\end{figure}

In equation (\ref{AC2}), if $\alpha G-k\omega _{0}=0$, the $k$-th current,
with the magnitude $I_{k}=\left\vert a\alpha J_{k}(\alpha g/\omega
_{0})/2bd^{\prime }\right\vert $, converts to a direct current. This means
that the Shapiro step, with the same magnitude $I_{k}$, emerges in our
predicted mass-current Josephson effect. We define a $k$-th gap $\Delta
_{k}=\alpha G_{0}-k\omega _{0}=N\omega _{0}$, where $N\in
%TCIMACRO{\U{2124} }%
%BeginExpansion
\mathbb{Z}
%EndExpansion
$ is a topological invariant of the fundamental group $\pi _{1}(S^{1})$ with
$S^{1}=\mathbf{\mathbb{%
%TCIMACRO{\U{211d} }%
%BeginExpansion
\mathbb{R}
%EndExpansion
}}^{1}\cup \{\infty \}$. We find that the direct-current component is
topologically trivial ($N=0$) and the alternating-current component is
topologically nontrivial ($N\neq 0$). When increasing $G$, we need consider
a generalized gap $\Delta _{k}=(N+\sigma )\omega _{0}$, where $\sigma \in
\lbrack 0,1]$. When $\sigma =1$, the $k$-th component becomes topologically
nontrivial ($N=1$), while the $(k+1)$-th component becomes topologically
trivial ($N=0$), where a new Shapiro step, with the magnitude $I_{k+1}$,
appears. This process is depicted in Fig. \ref{fig2}. We emphasize that our
predictions arises from the space-time-dependent non-Abelian gauge fields.
If the gauge field is chosen as the static Rashba-type gauge fields, $\Delta
\zeta _{\text{II\textbf{,}I}}\simeq d^{\prime }$, and only a constant
Josephson mass current emerges. This Josephson mass current is topologically
trivial, as shown in the black solid line in Fig. \ref{fig2}.

Finally, we briefly illustrate the possible experimental observation of the
predicted \textit{SU(2)} non-Abelian Josephson effect and the corresponding
Shapiro steps. In experiments, the double-well potential can be constructed
effectively by the superposition of a 1D periodic optical lattice with a 3D
magnetic harmonic trap. The frequencies in the radial and normal directions
of the 3D magnetic trap are of $10^{3}$Hz and $10^{2}$Hz, respectively \cite%
{LLAS04}. The width and height of barrier are about $2\sim 4\mu m$ and of $%
10^{3}$Hz, respectively. When the pairing field condenses in the double-well
potential with particle density $n=3\times 10^{13}cm^{-3}$, $G\sim 0.1E_{%
\text{F}}$ and $g\sim 0.05E_{\text{F}}$ \cite{MCC04}. This means the the
condition $\{G,g\}\ll E_{\text{F}}$ is valid. Thus, the predicted SU(2)
non-Abelian Josephson effect as well as the Shapiro steps can be detected
experimentally by the way of non-destructive phase contrast image \cite{LS07}%
. \newline

\bigskip

{\LARGE \textbf{Discussion}}

In summary, we have demonstrated strictly that the neutral pairing of
degenerate Fermi gases interacts with the same synthetic non-Abelian gauge
fields, imposing originally on the Fermi atoms. Moreover, we have obtained
the first and second \textit{SU(2)} GL equations, which allow us to predict
new quantum effects, such as an \textit{SU(2)} non-Abelian Josephson effect
and the corresponding Shapiro steps for the space-time-dependent non-Abelian
gauge fields. These results give new applications of the synthetic
non-Abelian gauge fields. For example, we can design a novel atomic
direct-current superconducting quantum interference device \cite{RCB13},
based on the predicted \textit{SU(2)} non-Abelian Josephson effect.

\bigskip

{\LARGE \textbf{Methods}}

\textbf{The local gauge theory of the pairing field. } In order to apply the
local gauge theory, we in this subsection consider the 4D
space-time-dependent coordinate, i.e., $x_{\mu }=(t,\mathbf{r})$. We begin
to study a two-component Fermi atom field $\phi (x_{\mu })$ coupled with the
synthetic non-Abelian gauge field. When the massive Fermi atom field
interacts with the synthetic non-Abelian gauge fields, its
behavior is identical to a Dirac field with the same local gauge symmetry.
In this Dirac-like atom field, each component reflects a spinor,
corresponding to an internal helical state. The corresponding speace-time
action is written as \cite{LR96}
\begin{equation}
S=\int d^{4}x_{\mu }[\bar{\phi}i\gamma ^{\mu }(\partial _{\mu }+i\alpha
A_{\mu })\phi -m\bar{\phi}\phi -\frac{1}{4}F_{\mu \nu }F^{\mu \nu }].
\label{TS}
\end{equation}%
In equation (\ref{TS}), $\gamma ^{\mu }$ ($\mu =0,1,2,3$) are the Dirac
gamma matrices, satisfying the Clifford algebra $\gamma ^{\mu }\gamma ^{\nu
}+\gamma ^{\nu }\gamma ^{\mu }=\delta _{\mu \nu }$\^{I}, where $\delta _{\mu
\nu }$ and \^{I} are the Kronecker notation and $4\times 4$ unit matrix,
respectively. $\partial _{\mu }+i\alpha A_{\mu }$ are the covariant
derivatives of the Fermi atom field $\phi $, where $A_{\mu }(x_{\mu })$ are
the synthetic non-Abelian gauge fields with $[A_{\mu },A_{\nu }]\neq 0$, and
$\alpha $ is a constant that governs the coupling between the Fermi atom
field $\phi $ and the non-Abelian gauge fields $A_{\mu }$. $F_{\mu \nu
}=\partial _{\mu }A_{\nu }-\partial _{\nu }A_{\mu }+i\alpha \lbrack A_{\mu
},A_{\nu }]$ is the tensor of our considered non-Abelian gauge fields. The
space-time action in equation (\ref{TS}) is invariant via a local gauge
transformation $\phi ^{\prime }=U_{L}\phi $, with $U_{L}(x_{\mu })=\exp
[-i\Lambda ^{\epsilon }(x_{\mu })\tau _{\epsilon }]$, where $\tau _{\epsilon
}$ ($\epsilon =1,2,3$) are the generators of the \textit{SU(2)} Lie group,
and $\Lambda ^{\epsilon }(x_{\mu })$ are the phase factors of space-time.

For the pairing field, we firstly investigate the global gauge symmetry, and
then generalize it to the local case. When we introduce a global \textit{%
SU(2)} operator $U_{G}=\exp (-i\Lambda _{G}^{a}\tau _{a})$, where $\Lambda
_{G}^{a}$ is independent of space-time, to make a gauge transformation $\phi
_{1}\rightarrow U_{G}\phi _{1}$ or $\phi _{2}\rightarrow U_{G}\phi _{2}$,
the pairing field becomes $\psi \rightarrow U_{G}\phi _{1}\phi _{2}$, which
means that $\psi ^{\prime }=U_{G}\psi $. According to the principle of
gauge-field theory, we should obtain a Lagrangian invariant $\mathcal{L}=T-V$%
, under the above global gauge transformation of the pairing field $\psi $.
Using the relation $-i\partial _{\mu }\psi ^{\prime }=-i\partial _{\mu
}U_{G}\psi =-iU_{G}\partial _{\mu }\psi $ and its complex conjugate, we find
directly that the kinetic energy $T=$ $(i\partial ^{\mu }\psi ^{\ast
})(-i\partial _{\mu }\psi )$ is invariant. For the scalar pairing field $%
\psi $ that can condense in a non-zero vacuum state, the effective potential
$V$ must have a stable and non-zero minimum point (vacuum). If expanding the
effective potential $V$ with respect to $\psi ^{\ast }\psi $ around the
critical temperature $T_{c}$ (up to second order), we obtain $V\simeq -a\psi
^{\ast }\psi -b(\psi ^{\ast }\psi )^{2}/2$. Thus, the global gauge-invariant
action for the pairing field $\psi $ is given by $S_{G}^{\psi }=\int
d^{4}X_{\mu }[(i\partial ^{\mu }\psi ^{\ast })(-i\partial _{\mu }\psi )+U_{%
\text{eff}}]$, where $U_{\text{eff}}=-V$ is an effective potential \cite%
{LR96}.

To discuss the local gauge symmetry of the pairing field $\psi $, we replace
$U_{G}$ by $U_{L}$ to make a similar gauge transformation. However, in such
case, $-i\partial _{\mu }\psi ^{\prime }\neq -iU_{L}\partial _{\mu }\psi $.
As a result, we introduce new covariant derivatives of the paring field $%
\psi $, $D_{\mu }=-i\partial _{\mu }-\beta B_{\mu }$, to realize $(D_{\mu
}\psi )^{\prime }=U_{L}(D_{\mu }\psi )$ \cite{LR96}, which gives rise to
three following equations:
\begin{equation}
-i\partial _{\mu }(U_{L}\psi )-\beta B_{\mu }^{\prime }U_{L}\psi
=-iU_{L}\partial _{\mu }\psi -\beta U_{L}B_{\mu }\psi ,\text{c.c.},
\label{LT1}
\end{equation}%
and
\begin{equation}
B_{\mu }^{\prime }=-\frac{i}{\beta }(\partial _{\mu }U_{L})U_{L}^{\dag
}+U_{L}B_{\mu }U_{L}^{\dag },  \label{LT2}
\end{equation}%
where c.c. is the complex conjugate. With the help of equation (\ref{LT2})
and the covariant derivatives $D_{\mu }$, we confirm that $(D^{\mu }\psi
^{\ast })(D_{\mu }\psi )$ are invariant under the local gauge transformation
$U_{L}$, and so is the effective potential $U_{\text{eff}}$. As a
consequence, we obtain the space-time action for the pairing field $\psi $
in the local gauge symmetry,
\begin{equation}
S_{L}^{\psi }=\int d^{4}X_{\mu }[(D^{\mu }\psi ^{\ast })(D_{\mu }\psi )+U_{%
\text{eff}}+\mathcal{L}_{V}],  \label{S2}
\end{equation}%
where $B_{\mu }$ are called the \textit{SU(2)} Yang-Mills gauge fields, $%
\beta $ is a constant reflecting the coupling between the pairing field $%
\psi $ and the Yang-Mills gauge fields $B_{\mu }$, $\mathcal{L}_{V}=-V_{\mu
\nu }V^{\mu \nu }/4$, with $V_{\mu \nu }=$ $D_{\mu }B_{\nu }-D_{\nu }B_{\mu
} $, is the energy density invariant of the Yang-Mills gauge fields $B_{\mu
} $.

Due to the identical gauge properties of the pairing field $\psi $ and the
Fermi atom field $\phi $, the Yang-Mills gauge fields $B_{\mu }$ must have
the same terms as the synthetic non-Abelian gauge fields $A_{\mu }$.
Moreover, they have an identical conserved quality called the \textit{SU(2)}
charge, according to Noether's theorem \cite{LR96}. This means that $\alpha
=\beta $. The above two results lead to a significant conclusion that the
pairing field $\psi $\textbf{\ }can\textbf{\ }also couple identically with
the non-Abelian gauge fields $A_{\mu }$, imposing originally on the Fermi
atoms, and have a similar internal helical doublet, like the Fermi atoms. In addition, we obtain equation (\ref{FEF}) in the text,
by extracting the spatial part of the space-time action in
equation (\ref{S2}).

\textbf{The derivation of the first and second \textit{SU(2)} GL equations.}
The variation of the total free energy functional can be written formally as
\begin{equation}
\delta F_{\text{s}}=\int d^{3}\mathbf{R(}\delta f_{\text{n}}+\delta U_{\text{%
eff}}+\delta f_{\text{c}}+\delta f_{\text{G}}).  \label{TV}
\end{equation}%
When condensate of the pairing field $\psi $ occurs, $\delta f_{\text{n}%
}\equiv 0$. Since the effective potential density does not depend on the
synthetic non-Abelian gauge fields, we have
\begin{equation}
\delta U_{\text{eff}}=\delta U_{\text{eff}}(\psi )+\delta U_{\text{eff}%
}(\psi ^{\ast }).  \label{dUeff}
\end{equation}%
If further neglecting the higher-order terms with respect to $\delta \psi $
and $\delta \psi ^{\ast }$, we derive $\delta U_{\text{eff}}(\psi )=a\delta
\psi \psi ^{\ast }+b\delta \psi \psi ^{\ast }\psi \psi $ and $\delta U_{%
\text{eff}}(\psi ^{\ast })=a\psi \delta \psi ^{\ast }+b\delta \psi ^{\ast
}\psi ^{\ast }\psi ^{\ast }\psi $.

For the coupled term between the pairing field $\psi $ and the synthetic
non-Abelian gauge fields $A_{i}$, we have
\begin{equation}
\delta f_{\text{c}}=\delta f_{\text{c}}(\psi )+\delta f_{\text{c}}(\psi
^{\ast })+\delta f_{\text{c}}(A_{i})\text{,}  \label{dFc}
\end{equation}%
where $i$ and $j$ run over $x$, $y$, and $z$, because the pairing has a 3D
momentum. After a careful calculation, we have $\delta f_{\text{c}}(\psi )=%
\left[ (\partial _{i}^{\ast }\psi ^{\ast }+i\alpha A_{i}\psi ^{\ast })\delta
\psi -(\partial _{i}^{\ast }+i\alpha A_{i})^{2}\psi ^{\ast }\delta \psi %
\right] /4m$ and $\delta f_{\text{c}}(\psi ^{\ast })=\left[ (\partial
_{i}\psi -i\alpha A_{i}\psi )\delta \psi ^{\ast }-(\partial _{i}-i\alpha
A_{i})^{2}\psi \delta \psi ^{\ast }\right] /4m$. On the other hand, when
neglecting the higher-order terms with respect to $\delta A_{i}$, we obtain $%
\delta f_{\text{c}}(A_{i})\simeq \alpha ^{2}\psi ^{\ast }\psi \delta
A_{i}A_{i}/2m+i\alpha (\psi ^{\ast }\partial _{i}\psi \delta A_{i}-\psi
\partial _{i}\psi ^{\ast }\delta A_{i})/4m$.

Finally, we consider the variation of the energy functional density of the
synthetic non-Abelian gauge fields $A_{i}$,
\begin{equation}
\delta f_{\text{G}}=\varrho \left[ L_{ij}(\mathbf{A+}\delta \mathbf{A}%
)L_{ij}(\mathbf{A+}\delta \mathbf{A})-L_{ij}(\mathbf{A})L_{ij}(\mathbf{A})%
\right] ,  \label{dfg}
\end{equation}%
where $L_{ij}(\mathbf{A})=-i\partial _{i}A_{j}+i\partial _{i}A_{j}+\alpha
A_{j}A_{i}-\alpha A_{i}A_{j}$ and $L_{ij}(\mathbf{A+}\delta \mathbf{A}%
)=-i\partial _{i}A_{j}+i\partial _{j}A_{i}-i\partial _{i}\delta
A_{j}+i\partial _{j}\delta A_{i}+\alpha A_{j}A_{i}-\alpha A_{i}A_{j}-\alpha
A_{i}\delta A_{j}+\alpha A_{j}\delta A_{i}-\alpha \delta A_{i}A_{j}+\alpha
\delta A_{j}A_{i}-\alpha \delta A_{i}\delta A_{j}+\alpha \delta A_{j}A_{i}$.
When neglecting all the high-order terms, such as $O^{2}(\delta A_{i})$, $%
O^{2}(\partial _{i}\delta A_{i})$, and $O^{2}(\delta A_{j}\partial
_{i}\delta A_{i})$, we obtain
\begin{eqnarray}
\delta f_{\text{G}} &=&2\varrho (\alpha A_{i}A_{j}-\alpha
A_{j}A_{i}+i\partial _{i}A_{j}-i\partial _{j}A_{i})  \label{DFG} \\
&&\times (\alpha A_{i}\delta A_{j}-\alpha A_{j}\delta A_{i}+i\partial
_{i}\delta A_{j}-i\partial _{j}\delta A_{i}  \notag \\
&&+\alpha \delta A_{i}A_{j}-\alpha \delta A_{j}A_{i})  \notag \\
&=&2\varrho (\Pi _{i}A_{j}-\Pi _{j}A_{i})(\Pi _{i}\delta A_{j}-\Pi
_{j}\delta A_{i})-  \notag \\
&&2\alpha \varrho (\Pi _{i}A_{j}-\Pi _{j}A_{i})(\delta A_{i}A_{j}-\delta
A_{j}A_{i}).  \notag
\end{eqnarray}%
Equation (\ref{DFG}) shows the properties induced by the synthetic
non-Abelian gauge fields $A_{i}$. If all non-commutators vanish, this
equation becomes $\delta f_{\text{G}}=-2\varrho (\partial _{i}A_{j}-\partial
_{j}A_{i})(\partial _{i}\delta A_{j}-\partial _{j}\delta A_{i})$, which is
the typical result for the Abelian gauge field in the \textit{U(1)} GL
theory.

In the presence of the Abelian gauge fields, we have $(\nabla \times \mathbf{%
A})\cdot (\nabla \times \delta \mathbf{A})=\delta \mathbf{A\cdot }(\nabla
\times \nabla \times \mathbf{A})-\nabla \cdot \lbrack (\nabla \times \mathbf{%
A})\times \delta \mathbf{A}]$. However, in the case of the \textit{SU(2)}
non-Abelian gauge fields only with the in-plane components (i.e., $A_{z}=0$%
), the above formula becomes $\partial _{\zeta }A_{\eta }\partial _{\zeta
}\delta A_{\eta }-\partial _{\zeta }A_{\eta }\partial _{\eta }\delta
A_{\zeta }-\partial _{\eta }A_{\zeta }\partial _{\zeta }\delta A_{\eta
}+\partial _{\eta }A_{\zeta }\partial _{\eta }\delta A_{\zeta }=(\partial
_{\eta }\partial _{\zeta }A_{\eta }-\partial _{z}^{2}A_{\zeta }-\partial
_{\eta }^{2}A_{\zeta })\delta A_{\zeta }+(\partial _{\eta }^{2}A_{\zeta
}\delta A_{\zeta }+\partial _{z}^{2}A_{\zeta }\delta A_{\zeta }+\partial
_{\zeta }A_{\eta }\partial _{\zeta }\delta A_{\eta }-\partial _{\zeta
}A_{\eta }\partial _{\eta }\delta A_{\zeta }+\partial _{z}A_{\eta }\partial
_{z}\delta A_{\eta }-\partial _{\zeta }\partial _{\eta }A_{\zeta }\delta
A_{\eta })$, and equation (\ref{DFG}) thus turns into
\begin{eqnarray}
\delta f_{\text{G}} &=&\varrho \lbrack -2(\partial _{\eta }\partial _{\zeta
}A_{\eta }-\partial _{z}^{2}A_{\zeta }-\partial _{\eta }^{2}A_{\zeta
})\delta A_{\zeta }+  \label{FGF} \\
&&4i\alpha \lbrack -(\partial _{\zeta }A_{\eta })A_{\eta }+(\partial _{\eta
}A_{\zeta })A_{\eta }]\delta A_{\zeta }  \notag \\
&&-2(\partial _{\eta }^{2}A_{\zeta }\delta A_{\zeta }+\partial
_{z}^{2}A_{\zeta }\delta A_{\zeta }+\partial _{\zeta }A_{\eta }\partial
_{\zeta }\delta A_{\eta }-  \notag \\
&&\partial _{\zeta }A_{\eta }\partial _{\eta }\delta A_{\zeta }+\partial
_{z}A_{\eta }\partial _{z}\delta A_{\eta }-\partial _{\zeta }\partial _{\eta
}A_{\zeta }\delta A_{\eta })].  \notag
\end{eqnarray}

In addition, for the 3D momentum of the pairing, the boundary conditions are
written as \cite{MC73}
\begin{equation}
(\partial _{i}^{\ast }+i\alpha A_{i})_{\mathbf{n}}\psi ^{\ast }=0,\text{c.c.}%
.  \label{BC}
\end{equation}%
Using these boundary conditions, the variation of the total free energy
functional is obtained by
\begin{equation}
\delta F_{\text{s}}=\delta F_{\text{s}}(\psi )+\delta F_{\text{s}}(\psi
^{\ast })+\delta F_{\text{s}}(A_{\zeta }),  \label{DFGF}
\end{equation}%
where
\begin{eqnarray}
\delta F_{\text{s}}(\psi ) &=&\int d^{3}\mathbf{R\{}\frac{1}{4m}[(i\partial
_{\zeta }^{\ast }+\alpha A_{\zeta })^{2}-\partial _{z}^{\ast 2}]\psi ^{\ast }
\label{R1} \\
&&+a\psi ^{\ast }+b\psi ^{\ast }\psi ^{2}\}\delta \psi ,  \notag
\end{eqnarray}%
\begin{eqnarray}
\delta F_{\text{s}}(\psi ^{\ast }) &=&\int d^{3}\mathbf{R\{}\frac{1}{4m}%
[(-i\partial _{\zeta }-\alpha A_{\zeta })^{2}-\partial _{z}^{2}]\psi
\label{R2} \\
&&+a\psi +b\psi ^{2}\psi \}\delta \psi ^{\ast },  \notag
\end{eqnarray}%
\begin{equation}
\begin{split}
\delta F_{\text{s}}(A_{\zeta })=& \frac{1}{2m}\int d^{3}\mathbf{R(}\alpha
^{2}\psi ^{\ast }\psi A_{\zeta } \\
& +\frac{i\alpha }{2}\mathbf{(}\psi ^{\ast }\partial _{\zeta }\psi -\psi
\partial _{\zeta }\psi ^{\ast }\mathbf{)}\delta A_{\mu } \\
& +\varrho \int d^{3}\mathbf{R}\{-2(\partial _{\eta }\partial _{\zeta
}A_{\eta }-\partial _{z}^{2}A_{\zeta }-\partial _{\eta }^{2}A_{\zeta
})\delta A_{\zeta } \\
& +4i\alpha \lbrack -(\partial _{\zeta }A_{\eta })A_{\eta }+(\partial _{\eta
}A_{\zeta })A_{\eta }]\delta A_{\zeta }\} \\
& +\varrho \int d^{3}\mathbf{R}[-2(\partial _{\eta }^{2}A_{\zeta }\delta
A_{\zeta }+\partial _{z}^{2}A_{\zeta }\delta A_{\zeta } \\
& +\partial _{\zeta }A_{\eta }\partial _{\zeta }\delta A_{\eta }-\partial
_{\zeta }A_{\eta }\partial _{\eta }\delta A_{\zeta }+\partial _{z}A_{\eta
}\partial _{z}\delta A_{\eta } \\
& -\partial _{\zeta }\partial _{\eta }A_{\zeta }\delta A_{\eta })].
\end{split}
\label{R3}
\end{equation}%
Finally, using the conditions $\delta F_{\text{s}}(\psi )=\delta F_{\text{s}%
}(\psi ^{\ast })=0$, we obtain the first GL equation (see equation (\ref{GL1}%
) in the text). In addition, by considering $\delta F_{\text{s}}(A_{\zeta
})=0$, we derive the second GL equation and the supercurrents in the $x$, $y$%
, and $z$ directions (see equations (\ref{GL2})-(\ref{R8}) in the text).

\textbf{The derivation of equation (\ref{AC2}).} We rewrite equation (\ref%
{AC1}) as
\begin{equation}
j_{\zeta }^{m}=-\frac{a\alpha }{2bd^{\prime }}\hat{\Xi}\mbox{Im}\lbrack \exp
(\left[
\begin{array}{cc}
0 & \varpi _{1} \\
\varpi _{2} & 0%
\end{array}%
\right] )],  \label{JM1}
\end{equation}%
where $\varpi _{1}=-i\alpha \lambda _{2}[Gt+g\sin (\omega _{0}t)/\omega
_{0}+\Delta \varphi _{01}]$ and $\varpi _{2}=-i\alpha \lambda _{1}[Gt+g\sin
(\omega _{0}t)/\omega _{0}+\Delta \varphi _{02}]$. When $\{G,g\}\ll E_{\text{%
F}}$, we approximately obtain
\begin{equation}
\begin{split}
j_{\zeta }^{m}=& -\frac{a\alpha }{2bd^{\prime }}\hat{\Xi} \\
\times & \mbox{Im}\lbrack \exp (\left[
\begin{array}{cc}
0 & \frac{\alpha \lambda _{1}}{i}(Gt+\Delta \varphi _{01}) \\
\frac{\alpha \lambda _{2}}{i}(Gt+\Delta \varphi _{02}) & 0%
\end{array}%
\right] ) \\
& \times \exp (\left[
\begin{array}{cc}
0 & \frac{\alpha g\lambda _{1}}{i\omega _{0}}\sin (\omega _{0}t) \\
\frac{\alpha g\lambda _{2}}{i\omega _{0}}\sin (\omega _{0}t) & 0%
\end{array}%
\right] )].
\end{split}
\label{JM2}
\end{equation}%
Based on the definition of matrix exponential, equation (\ref{JM2}) turns
into
\begin{equation}
j_{\zeta }^{m}=-\frac{a\alpha }{2bd^{\prime }}\hat{\Xi}\mathbf{\mbox{Im}(}%
\hat{P}\mathbf{),}  \label{JM3}
\end{equation}%
where $\hat{P}=$diag$(\exp [i\alpha (Gt+\Delta \varphi _{01})]\exp [i\alpha
g\sin (\omega _{0}t)/\omega _{0}]\lambda _{1}\lambda _{2}$, $\exp [i\alpha
(Gt+\Delta \varphi _{02})]\exp [i\alpha g\sin (\omega _{0}t)/\omega
_{0}]\lambda _{1}\lambda _{2})$. To analyze the properties of the gauge
invariant mass current, we need take a Fourier-Bessel power series for the
elements of the matrix $\hat{P}$. By considering the parity of the Bessel
function, i.e., $J_{k}(x)=(-1)^{k}J_{-k}(x)$, we have
\begin{equation}
\begin{split}
\exp [i\frac{\alpha g}{\omega _{0}}& \sin \left( \omega _{0}t\right) ]= \\
=& \sum\limits_{k=-\infty }^{\infty }J_{k}(\frac{\alpha g}{\omega _{0}})\cos
(k\omega _{0}t)+iJ_{k}(\frac{\alpha g}{\omega _{0}})\sin (k\omega _{0}t) \\
=& \sum\limits_{k=-\infty }^{\infty }(-1)^{k}J_{k}(\frac{\alpha g}{\omega
_{0}})\exp (-ik\omega _{0}t).
\end{split}
\label{FBS}
\end{equation}%
Substitute equation (\ref{FBS}) into the matrix $\hat{P}$ yields equation (%
\ref{AC2}).\newline

\bigskip {\textbf{Acknowledgements}} We thank Professors Ming Gong, An-chun
Ji, and Qing Sun for their valuable discussions. This work is supported
partly by the 973 program under Grant No. 2012CB921603; the NNSFC under
Grant No. 61275211; the PCSIRT under Grant No. IRT13076; the NCET under
Grant No. 13-0882; the FANEDD under Grant No. 201316; the OIT under Grant
No. 2013804; and OYTPSP. C.Z. is supported partly by ARO (W911NF-12-1-0334),
AFOSR (FA9550-13-1-0045), and NSF-PHY (1249293).

{\textbf{Author Contributions}} C.Z., G.C., and S.J. conceived the idea,
K.Z., Y.F., and G.C. performed the calculations, C.Z., G.C., and S.J. wrote
the manuscript and supervised the whole research project. Correspondence and
requests for materials should be addressed to G.C. (chengang971@163.com).

{\textbf{Competing Interests}} The authors declare that they have no
competing financial interests.

{\textbf{Author Information}} Reprints and permissions information is
available at www.nature.com/reprints.


\begin{thebibliography}{99}
\bibitem{RJ95} Ruseckas, J., Juzeli\={u}nas, G., \"{O}hberg, P. \&
Fleischhauer, M. Non-Abelian gauge potentials for ultracold atoms with
degenerate dark states. Phys. Rev. Lett. \textbf{95}, 010404 (2005).

\bibitem{JD11} Dalibard, J., Gerbier, F., Juzeli\={u}nas, G. \& \"{O}hberg,
P. Artificial gauge potentials for neutral atoms. Rev. Mod. Phys. \textbf{83}%
, 1523-1543 (2011).

\bibitem{NG14} Goldman, N., Juzeli\={u}nas, G., \"{O}hberg, P. \& Spielman
I. B. Light-induced gauge fields for ultracold atoms. arXiv: 1308.6533.

\bibitem{YJL11} Lin, Y.-J., Jim\'{e}nez-Garc\'{\i}a, K. \& Spielman, I. B.
Spin-orbit-coupled Bose-Einstein condensates. Nature \textbf{471}, 83-86
(2011).

\bibitem{JYZ12} Zhang, J. Y. \textit{et al.} Collective dipole oscillations
of a spin-orbit coupled Bose-Einstein condensate. Phys. Rev. Lett. \textbf{%
109}, 115301 (2012).

\bibitem{CQ13} Qu, C., Hamner, C., Gong, M., Zhang, C. \& Engels, P.
Observation of Zitterbewegung in a spin-orbit-coupled Bose-Einstein
condensate. Phys. Rev. A \textbf{88}, 021604 (2013).

\bibitem{JSC14} Ji, S.-C. \textit{et al.} Experimental determination of the
finite-temperature phase diagram of a spin-orbit coupled Bose gas. Nat.
Phys. \textbf{10}, 314-320 (2014).

\bibitem{CH14} Chris, H. \textit{et al.} Dicke-type phase transition in a
spin-orbit coupled Bose-Einstein condensate. Nat. Commun. in press (2014).

\bibitem{PW12} Wang, P. \textit{et al.} Spin-orbit coupled degenerate Fermi
gases. Phys. Rev. Lett. \textbf{109}, 095301 (2012).

\bibitem{RAW13} Williams, R. A., Beeler, M. C., LeBlanc, L. J., Jim\'{e}%
nez-Garc\'{\i}a, K. \& Spielman, I. B. Raman-induced interactions in a
single-component Fermi gas near an s-wave Feshbach resonance. Phys. Rev.
Lett. \textbf{111}, 095301 (2013).

\bibitem{ZF14} Fu, Z. \textit{et al.} Production of Feshbach molecules
induced by spin-orbit coupling in Fermi gases. Nat. Phys. \textbf{10},
110-115 (2014).

\bibitem{LWC12} Cheuk, L. W. \textit{et al.} Spin-injection spectroscopy of
a spin-orbit coupled Fermi gas. Phys. Rev. Lett. \textbf{109}, 095302 (2012).

\bibitem{MCB13} Beeler, M. C. \textit{et al.} The spin Hall effect in a
quantum gas. Nature \textbf{498}, 201-204 (2013).

\bibitem{YZQ11} Yu, Z.-Q. \& Zhai, H. Spin-orbit coupled Fermi gases across
a Feshbach resonance. Phys. Rev. Lett. \textbf{107}, 195305 (2011).

\bibitem{HH11} Hu, H., Jiang, L., Liu, X.-J. \& Pu, H. Probing anisotropic
superfluidity in atomic Fermi gases with Rashba spin-orbit coupling. Phys.
Rev. Lett. \textbf{107}, 195304 (2011).

\bibitem{VJP11} Vyasanakere, J. P., Zhang, S. \& Shenoy V. B. BCS-BEC
crossover induced by a synthetic non-Abelian gauge field. Phys. Rev. B
\textbf{84}, 014512 (2011).

\bibitem{HL12} He, L. \& Huang, X.-G. BCS-BEC Crossover in 2D Fermi gases
with Rashba spin-orbit coupling. Phys. Rev. Lett. \textbf{108}, 145302
(2012).

\bibitem{RYW12} Liao, R., Yu, Y.-X. \& Liu, W.-M. Tuning the tricritical
point with spin-orbit coupling in polarized Fermionic condensates. Phys.
Rev. Lett. \textbf{108}, 080406 (2012).

\bibitem{WF13L} Wu, F., Gou, G.-C., Zhang, W., \& Yi, W. Unconventional
superfluid in a two-dimensional Fermi gas with anisotropic spin-orbit
coupling and Zeeman fields. Phys. Rev. Lett. \textbf{110}, 110401 (2013).

\bibitem{WF13A} Wu, F., Gou, G.-C., Zhang, W., \& Yi, W. Unconventional
Fulde-Ferrell-Larkin-Ovchinnikov pairing states in a Fermi gas with
spin-orbit coupling. Phys. Rev. A \textbf{88}, 043614 (2013).

\bibitem{YX14} Xu, Y., Chu, R.-L. \& Zhang, C., Anisotropic Weyl fermions
from the quasiparticle excitation spectrum of a 3D Fulde-Ferrell superfluid.
Phys. Rev. Lett. \textbf{112}, 136402 (2014).

\bibitem{MG11} Gong, M., Tewari, S. \& Zhang, C. BCS-BEC crossover and
topological phase transition in 3D spin-orbit coupled degenerate Fermi
gases. Phys. Rev. Lett. \textbf{107}, 195303 (2011).

\bibitem{MG12} Gong, M., Chen, G., Jia, S. \& Zhang, C. Searching for
Majorana Fermions in 2D spin-orbit coupled Fermi superfluids at finite
temperature. Phys. Rev. Lett. \textbf{109}, 105302 (2012).

\bibitem{KS12} Seo, K., Han, L. \& S\'{a} de Melo, C. A. R. Emergence of
Majorana and Dirac particles in ultracold Fermions via tunable interactions,
spin-orbit effects, and Zeeman fields. Phys. Rev. Lett. \textbf{109}, 105303
(2012).

\bibitem{HH13} Hu, H., Jiang, L., Pu, H., Chen, Y. \& Liu, X.-J. Universal
impurity-induced bound state in topological superfluids. Phys. Rev. Lett.
\textbf{110}, 020401 (2013).

\bibitem{CC13} Chen, C. Inhomogeneous topological superfluidity in
one-dimensional spin-orbit-coupled Fermi gases. Phys. Rev. Lett. \textbf{111}%
, 235302 (2013).

\bibitem{CQU13} Qu, C. \textit{et al.} Topological superfluids with
finite-momentum pairing and Majorana fermions. Nat. Commun. \textbf{4}, 2710
(2013).

\bibitem{WZ13} Zhang, W. \& Yi, W. Topological
Fulde-Ferrell-Larkin-Ovchinnikov states in spin-orbit-coupled Fermi gases.
Nat. Commun. \textbf{4}, 3710 (2013).

\bibitem{XJL13} Liu, X.-J. \& Hu, H. Inhomogeneous topological superfluidity
in one-dimensional spin-orbit-coupled Fermi gases. Phys. Rev. A \textbf{88},
023622 (2013).

\bibitem{CCF14} Chan, C. F. \& Gong, M. Pairing symmetry, phase diagram, and
edge modes in the topological Fulde-Ferrell-Larkin-Ovchinnikov phase. Phys.
Rev. B \textbf{89}, 174501 (2014).

\bibitem{HH14} Hu, H., Dong, L., Cao, Y., Pu, H. \& Liu, X.-J. Gapless
topological Fulde-Ferrell superfluidity induced by in-plane Zeeman field.
arXiv: 1404.2442.

\bibitem{MC73} Cyrot, M. Ginzburg-Landau theory for superconductors. Rep.
Prog. Phys. \textbf{36}, 103 (1973).

\bibitem{CAR93} S\'{a} de Melo, C. A. R., Randeria, M. \& Engelbrecht, J. R.
Crossover from BCS to Bose superconductivity: transition temperature and
time-dependent Ginzburg-Landau theory, Phys. Rev. Lett. \textbf{71},
3202-3205 (1993).

\bibitem{JFA04} James, F. A. \textit{Superconductivity, Superfluids, and
Condensates} (Oxford University Press, New York, 2004).

\bibitem{AJL06} Leggett, A. J. \textit{Quantum Liquids: Bose Condensation
and Cooper Pairing in Condensed-Matter Systems} (Oxford University Press,
New York, 2006).

\bibitem{AAA57} Abrikosov, A. A. On the magnetic properties of
superconductors of the second group, Sov. Phys. JETP \textbf{5, }1174-1182
(1957).

\bibitem{BR10} Rosenstein, B. \& Li, D. Ginzburg-Landau theory of type II
superconductors in magnetic field, Rev. Mod. Phys. \textbf{82}, 109-168
(2010).

\bibitem{SW67} Weinberg, S. A model of leptons, Phys. Rev. Lett. \textbf{19}%
, 1264-1266 (1967).

\bibitem{LR96} Lewish, R. \textit{Quantum Field Theory} (Cambridge
University Press, Cambridge, 1996).

\bibitem{BD65} Josephson, B. D. Supercurrents though barriers. Adv. Phys.
\textbf{14}, 419-451 (1965).

\bibitem{LS07} Levy, S., Lahoud, E., Shomrni, I. \& Steinhauer, J. The a.c.
and d.c. josephson effects in a Bose-Einstein condensate. Nature \textbf{449}%
, 579-583 (2007).

\bibitem{AM05} Albiez, M. \textit{et al.} Direct observation of tunneling
and nonlinear self-trapping in a single bosonic Josephson junction. Phys.
Rev. Lett. \textbf{95}, 010402 (2005).

\bibitem{SA07} Spuntarelli, A., Pieri, P. \& Strinati, G. C. Josephson
effect thoughout the BCS-BEC crossover. Phys. Rev. Lett. \textbf{99}, 040401
(2007).

\bibitem{HH111} Hu, H. \& Liu, X.-J. Josephson effect in atomic
Fulde-Ferrell-Larkin-Ovchinnikov superfluid. Phys. Rev. A \textbf{83},
013631 (2011).

\bibitem{BPA98} Anderson, B. P. \& Kasevich, M. A. Macroscopic quamtum
interference from atomic tunnel arrys. Science \textbf{282}, 1686-1689
(1998).

\bibitem{FSC01} Cataliotti, F. S. \textit{et al.} Josephson junction arrays
with Bose-Einstein condensates. Science \textbf{293}, 843-846 (2001).

\bibitem{LLAS04} Pezz\`{e}, L. \textit{et al.} Insulating behavior of
trapped ideal fermi gas. Phys. Rev. Lett. \textbf{93}, 120401 (2004).

\bibitem{MCC04} Zwierlein, M. W. \textit{et al}. Condensation of pairs of
fermionic atoms near a Feshbach resonance. Phys. Rev. Lett. \textbf{92},
120403 (2004).

\bibitem{RCB13} Ryu, C., Blackburn, P. W., Blinova, A. A. \& Boshier, M. G.
Experimental realization of Josephson junctions for an atom SQUID. Phys.
Rev. Lett. \textbf{111}, 205301 (2013).
\end{thebibliography}
\end{document}